\documentclass[aps,pra,superscriptaddress,twocolumn,showpacs]{revtex4}

\usepackage{amsmath}
\usepackage{amssymb}
\usepackage{graphicx}
\usepackage{dcolumn}
\usepackage{bm}

\begin{document}

\title{Phase transition to Bose-Einstein condensation for a Bosonic gas confined in a combined trap}

\author{Baolong L\"{u}}
\affiliation{State Key Laboratory of Magnetic Resonance and Atomic
and Molecular Physics, Wuhan Institute of Physics and Mathematics,
Chinese Academy of Sciences, Wuhan 430071, People's Republic of
China}

\author{Xinzhou Tan}
\affiliation{State Key Laboratory of Magnetic Resonance and Atomic
and Molecular Physics, Wuhan Institute of Physics and Mathematics,
Chinese Academy of Sciences, Wuhan 430071, People's Republic of
China} \affiliation{Graduate School of the Chinese Academy of
Sciences, Beijing 100049, People's Republic of China}

\author{Bing Wang}
\affiliation{State Key Laboratory of Magnetic Resonance and Atomic
and Molecular Physics, Wuhan Institute of Physics and Mathematics,
Chinese Academy of Sciences, Wuhan 430071, People's Republic of
China} \affiliation{Graduate School of the Chinese Academy of
Sciences, Beijing 100049, People's Republic of China}

\author{Lijuan Cao}
\affiliation{State Key Laboratory of Magnetic Resonance and Atomic
and Molecular Physics, Wuhan Institute of Physics and Mathematics,
Chinese Academy of Sciences, Wuhan 430071, People's Republic of
China} \affiliation{Graduate School of the Chinese Academy of
Sciences, Beijing 100049, People's Republic of China}

\author{Hongwei Xiong}
\email{xionghongwei@wipm.ac.cn}\affiliation{State Key Laboratory of
Magnetic Resonance and Atomic and Molecular Physics, Wuhan Institute
of Physics and Mathematics, Chinese Academy of Sciences, Wuhan
430071, People's Republic of China}

\date{\today }

\begin{abstract}

We present a study of phase transition to macroscopic superfluidity
for an ultracold bosonic gas confined in a combined trap formed by a
one-dimensional optical lattice and a harmonic potential, focusing
on the critical temperature of this system and the interference
patterns of the Bose gas released from the combined trap. Based on a
semiclassical energy spectrum, we develop an analytic approximation
for the critical temperature $T_{c}$, and compare the analytic
results with that obtained by numerical computations. For finite
temperatures below $T_{c}$, we calculate the interference patterns
for both the normal gas and the superfluid gas. The total
interference pattern shows a feature of ``peak-on-a-peak." As a
comparison, we also present the experimentally observed interference
patterns of $^{87}$Rb atoms released from a one-dimensional optical
lattice system in accord with our theoretical model. Our
observations are consistent with the theoretical results.

\end{abstract}

\begin{pacs}
  {03.75.Lm, 67.25.dj, 37.10.Jk, 67.10.Ba}
\end{pacs}

\maketitle

\section{INTRODUCTION}
\label{Introduction}

Bosonic atoms confined in optical lattices have proved to be a
unique laboratory for investigating quantum phase transitions from
superfluids to Mott insulators \cite{U-Greiner,Jaksch,Stoferle}. The
momentum distribution of a lattice system can be mapped out directly
by the interference pattern of the atomic cloud after a ballistic
expansion over a time of flight (TOF). The emergence of macroscopic
bosonic superfluid is usually identified by the appearance of
interference peaks. However, recent theoretical works \cite{Ho,Kato}
for homogeneous gases in a three-dimensional (3D) lattice showed
that this criterion of macroscopic superfluidity is not reliable
since even a normal gas can have sharp interference peaks. The
underlying physical picture is that a lattice system at finite
temperatures possesses a ``V-shaped" phase diagram
\cite{Ho,Kato,Cha} which includes a normal gas region between the
Mott Insulator and the superfluid. The true signature of macroscopic
superfluidity is the $\delta$-function momentum peaks with nearly
unit visibility \cite{Ho}. Below critical temperature, the
coexistence of superfluid and normal gas in the homogeneous lattice
system should give rise to an interference pattern having a feature
of ``peak on a peak" \cite{Kato}. The new criterion of macroscopic
superfluidity makes it necessary to further investigate the phase
transition of bosonic atoms in an optical lattice, particularly for
the characteristics associated with the critical temperature and
interference patterns. Experimental investigations are also required
for comparison with relevant theoretical models.

There have been a few theoretical works
\cite{Kleinert,Ramakumar,Zobay} considering the translationally
invariant (uniform) lattices. However, in a realistic experiment, an
optical lattice is always accompanied by harmonic confinement in all
dimensions, arising from the focused Gaussian laser beams and/or an
external magnetic trap. A bosonic gas is, therefore, never spatially
uniform over the lattice range. Wild \emph{et al}. \cite{Tc-Wild}
have examined the critical temperature of the interacting bosons in
a one-dimensional (1D) lattice with additional harmonic confinement.
Ramakumar \emph{et al}. \cite{Tc of 123D-Ramakumar} have
investigated the condensate fraction and specific heat of
non-interacting bosons in 1D, two-dimensional (2D), and 3D lattices
in the presence of harmonic potentials. Based on a piecewise
analytic density of states extended to excited bands, Blakie
\emph{et al}. \cite{Tc-Blakie} developed an analytical expression of
the critical temperature for an ideal bosonic gas in the combined
harmonic lattice potential, and compared the analytic result with
their numerical computations. However, these studies on combined
traps did not mention interference patterns of the released bosonic
gases. A more recent theoretical paper \cite{Duan} has investigated
the Bose-Einstein condensation (BEC) in a 3D inhomogeneous optical
lattice system, and predicted that a bimodal structure in the
momentum-space density profile is a universal indicator of BEC
transition.

The experiment of Spielman \emph{et al}. \cite{Phillips} has
examined the superfluid to normal transition for a finite-sized 2D
optical lattice system. Their measurements confirm that bimodal
momentum distributions are associated with the superfluid phase. For
such a system with a typical density of $1$ atom per lattice site,
the phase transition behavior can be interpreted in terms of the
commonly used Bose-Hubbard model.

Unlike the 2D and 3D cases, an inhomogeneous 1D optical lattice
system is usually much more heavily populated, with an atomic number
up to several hundreds in a single lattice site. In the superfluid
phase, the on-site interaction energy $U$ varies from site to site
because of its dependence on the local population in single lattice
sites. This increases the complexity in searching for an analytical
description of the phase transition. In this paper we present a
study of the critical temperature and interference patterns of an
ultracold bosonic gas confined by a 1D optical lattice and an
additional magnetic potential. The interference patterns of the
normal gas and the condensed atoms are treated separately. The
superposition of the two parts gives rise to a feature of ``peak on
a peak." Different from a homogeneous lattice system, however, the
normal gas can-not produce sharp interference peaks. Furthermore,
the theoretical results are compared with our preliminary experiment
for a 1D lattice system of $^{87}$Rb Bose-Einstein condensates.

Our theoretical model relevant to the phase transition is for ideal bosonic
gases. In fact, interatomic interaction may affect the shape of the
interference pattern, especially for the condensed part which has a higher
atomic density. In order to obtain a better match with the experiment result,
we take the interaction energy into consideration for the condensed atoms
during the time of flight. The computed result shows that interference peaks
arising from the condensed atoms can be significantly broadened due to the
interaction effect.

This paper is organized as follows. In Sec. \ref{Critical
Temperature}, we begin with a semiclassical energy spectrum for a
combined harmonic lattice trap. Under the tight-binding
approximation and in the low energy limit, we derive an analytical
expression of the critical temperature for the atoms condensed to a
superfluid state. The accuracy of the analytical results are checked
with respect to the numerical calculations. Section \ref{Peaks}
gives a description on how the interference patterns are calculated
for the normal gas, as well as the Bose-condensed gas. In Sec.
\ref{Experiment}, we briefly introduce the experiment, and present
the observed interference patterns for a comparison with our
theoretical results. Finally in Sec. \ref{Conclusion}, we summarize
the obtained results.

\section{CRITICAL TEMPERATURE}
\label{Critical Temperature}

We now consider an ideal Bose gas confined in a 3D harmonic
potential with axial symmetry around the $z$ direction. The axial
and transverse trapping frequencies are $\omega_{z}$ and
$\omega_{x}=\omega_{y}=\omega_{\bot}$, respectively. Moreover, we
assume that the axial confinement is much weaker than the radial
confinement ($\omega_{z}\ll\omega_{\bot}$), so that the Bose gas is
made cigar-shaped. A 1D optical lattice potential,
$V_{0}\sin^{2}(kz)$, is applied along the $z$ axis, where $k=\pi/d$
is the wave vector of the lattice light, $d$ denotes the lattice
period, and $V_{0}$ denotes the potential depth of the lattice.
$V_{0}$ can be written in terms of the recoil energy $E_{r}$, say,
$V_{0}=sE_{r}$, where $E_{r}=\hbar^{2}k^{2}/2m$, and $m$ is the
atomic mass. The harmonic potential, together with the optical
lattice, forms a combined trap written as
\begin{equation}\label{Combined trap}
V(x,y,z)=\frac{1}{2}m\omega_{\bot}^{2}(x^{2}+y^{2})+\frac{1}{2}m\omega_{z}^{2}z^{2}+V_{0}\sin^{2}(kz).
\end{equation}
In practice, an optical lattice is usually produced by a retro-reflected
Gaussian laser beam which also produces a transverse confining potential, that
can be simply absorbed into $\omega_{\bot}$ if it is non-negligible.

To obtain the eigenenergies of the combined trap system, one needs
to derive the single-particle Hamiltonian of the system and then
numerically diagonalize it \cite{Tc-Blakie}. Despite its accuracy
for ideal Bose gases, this numerical method can not provide an
analytic expression of the energy levels. The energy spectrum
corresponding to the transverse confinement is described by equally
spaced harmonic-oscillator states, whereas the oscillator treatment
is not applicable to the axial dimension due to the presence of the
optical lattice. Our discussion hereafter is based on the
tight-binding approximation that only the ground band is accessible
to the system. This approximation is valid when the thermal energy
of the atoms is much less than the first band gap of a deep lattice.
For a 1D uniform lattice, the eigen energy can be written as a
function of quasimomentum $q$ \cite{J-Zwerger},
$\epsilon(q)=\frac{1}{2}\hbar\widetilde{\omega}-2J\cos(qd/\hbar)$.
Here, $\widetilde{\omega}$ is the frequency of the local oscillation
at each lattice well, while $J$ is the tunneling energy due to the
hopping to a nearest neighboring well, and it depends upon the
lattice depth $s$ in the following form \cite{J-Zwerger}
\begin{equation}\label{J}
J=\frac{4}{\sqrt{\pi}}E_{r}s^{3/4}\exp\left(-2\sqrt{s}\right).
\end{equation}
It should be noted that Eq. \eqref{J} is valid only for deep lattices. At
$s=11$, for example, it overestimates $J$ by approximately $18\%$. For the
combined trap, it is a reasonable assumption that Eq. \eqref{J} remains valid
as long as the trapping frequency $\omega_{z}$ of the weak axial confinement
is much smaller than the tunneling rate $J/\hbar$. We are thus able to use a
constant $J$ over the entire lattice system at a given lattice depth. For
simplicity the energy spectrum corresponding to the combined confinement in
the axial direction is approximated by the semiclassical energy,
$\epsilon(p_{z})+\frac{1}{2}m\omega_{z}^{2}z^{2}$, in the $z$\--\-$p_{z}$
phase space, where $p_{z}$ is the quasimomentum in the ground band. Now we are
able to write the total energy spectrum in an explicit form
\begin{equation}\label{energy}
\begin{split}
\varepsilon_{n_{x}n_{y}}(z,p_{z})=& \hbar\omega_{\bot}(n_{x}+n_{y}+1)
+\frac{1}{2}m\omega_{z}^{2}z^{2}\\
&+\frac{1}{2}\hbar\widetilde{\omega}_{z}-2J\cos(p_{z}d/\hbar),
\end{split}
\end{equation}
where $\{n_{x},n_{y}\}$ are non-negative integers.

For a semiclassical description of this system, we treat the harmonic trap
semiclassically while treating the optical lattice quantum mechanically. Such
a picture corresponds to a density distribution of the thermal cloud:
\begin{equation}\label{N-z-semi}
n(z)=\sum_{n_{x},n_{y}}\int\frac{\text{d}p_{z}}{2\pi\hbar}
F(p_{z},z)Md|\Phi_{p_{z}}(z)|^{2} ,
\end{equation}
where
\begin{equation*}\label{F-pz-z}
F(p_{z},z)=\frac{1}{\exp[\beta(\varepsilon_{n_{x}n_{y}}-\mu)]-1} ,
\end{equation*}
and
\begin{equation*}\label{Phipz-z}
\Phi_{p_{z}}(z)=\frac{1}{\sqrt{M}}\sum_{l=-M/2}^{M/2}w(z-ld)\exp\left(ip_{z}z/\hbar\right).
\end{equation*}
Here, $\Phi_{p_{z}}$ is the normalized wave function of a uniform optical
lattice system with an extension of $M$ lattice sites, and $w(z-ld)$ is the
Wannier wave function. The total number of thermal atoms is then written as
\begin{equation}\label{Nthermal-A}
\begin{split}
N_{th}&=\int n(z)\text{d}z  \\
&=\sum_{n_{x},n_{y}}\int\frac{\text{d}p_{z}\text{d}z}{2\pi\hbar}
F(p_{z},z)Md|\Phi_{p_{z}}(z)|^{2}.
\end{split}
\end{equation}
In the tight-binding limit, $w(z)$ is well localized within a single lattice
site. In contrast, $F(p_{z},z)$ is a slowly varying function of $z$.
Therefore, $\Phi_{p_{z}}(z)$ in Eq. \eqref{Nthermal-A} can be integrated out.
This results in a new integrand expressed as a summation of discrete
$F(p_{z},z-ld)d$, which in turn can be approximated as an integral over $z$.
By doing so, one gets
\begin{equation}\label{Nthermal-B}
N_{th}=\sum_{n_{x},n_{y}}\int\frac{\text{d}p_{z}\text{d}z}{2\pi\hbar}
F(p_{z},z).
\end{equation}

Below a critical temperature $T_{c}$ the chemical potential $\mu$ of the Bose
gas reaches the bottom of the ground band
\begin{equation*}\label{mu-c}
\mu\rightarrow\mu_{c}=\hbar\omega_{\bot}+\frac{1}{2}\hbar\widetilde{\omega}_{z}-2J,
\end{equation*}
while the lowest state with $p_{z}=0$ becomes macroscopically
populated which corresponds to the onset of Bose-Einstein
condensation. The condensed atoms exhibit macroscopic superfluidity,
whereas all other atoms beyond the lowest state form a so-called
normal gas. Since the condensate is actually a quantum fluid, we use
``superfluid" just as a synonym of BEC. The atomic number of the
normal gas is given by the sum,
\begin{equation}\label{N-nc}
N_{nc}=N-N_{c}=\sum_{n_{x},n_{y}}\int\frac{1}{\exp[\beta(\varepsilon_{n_{x}n_{y}}-\mu_{c})]-1}
\frac{\text{d}p_{z}\text{d}z}{2\pi\hbar},
\end{equation}
where $N$ is the total number of the atoms, $N_{c}$ the atomic number of the
condensed part, $\beta=1/k_{B}T$, and $k_{B}$ the Boltzmann constant. The
integrand can be expanded in powers of the exponential term using the formula
$1/(e^{x}-1)=\sum_{n=1}^{\infty}e^{-nx}$. Moreover, the sum over $n_{x}$,
$n_{y}$ can be replaced by an integral if the atomic number $N$ is large.
Performing the integration over $n_{x}$, $n_{y}$ as well as that over the
coordinate $z$, one gets
\begin{equation}\label{N-nc-pz}
\begin{split}
N-N_{c}=&
\frac{1}{(\beta\hbar\omega_{\bot})^{2}}\sum_{n=1}^{\infty}\frac{1}{n^{2}}\left(\frac{2\pi}{n\beta
m\omega_{z}^{2}}\right)^{1/2}\\
 & \times\int\frac{\text{d}p_{z}}{2\pi\hbar}\exp[-n\beta(2J-2J\cos(p_{z}d/\hbar))],
\end{split}
\end{equation}
and the right side is a function of temperature $T$. Apparently, Eq.
\eqref{N-nc-pz} is suitable for numerical calculation of the atomic number in
the normal gas since the integration can be simply replaced by a summation
over the $p_{z}$ region.

By imposing that $N_{c}=0$ at the transition, Eq. \eqref{N-nc-pz} determines a
critical temperature $T_{c}$ for a given $N$. Apparently, to obtain the value
of $T_{c}$, one needs to carry out numerical computations based on Eq.
\eqref{N-nc-pz}. Nevertheless, we can derive an analytic expression of $T_{c}$
in a limiting case. When the temperature of the Bose gas is so low that most
atoms occupy the states in the vicinity of the bottom of the ground band, the
relation $p_{z}\ll \hbar/d$ holds, and the cosine function in Eq.
\eqref{N-nc-pz} can be expanded to the order of $p_{z}^{2}$. With the
$p_{z}$\--\-dependent function integrated out, one has
\begin{equation*}\label{Nnc-mstar}
N=\left(k_{B}T_{c}/\hbar
\overline{\omega}\right)^{3}\left(m^{*}/m\right)^{1/2}\zeta(3),
\end{equation*}
where $\overline{\omega}=(\omega_{\bot}^{2}\omega_{z})^{1/3}$ is the geometric
average of the trapping frequencies, $m^{*}=\hbar^{2}/2Jd^{2}$ the effective
mass of the atom, and $\zeta(\alpha)=\sum_{n=1}^{\infty}1/n^{\alpha}$ the
Riemann zeta function. Finally, one gets
\begin{equation}\label{Tc}
k_{B}T_{c}=0.94\hbar\overline{\omega}N^{1/3}\left(m/m^{*}\right)^{1/6},
\end{equation}
which can be used as an analytic estimation of the critical temperature.
\begin{figure}[h]
\centering
\includegraphics[width=0.7\columnwidth,angle=0]{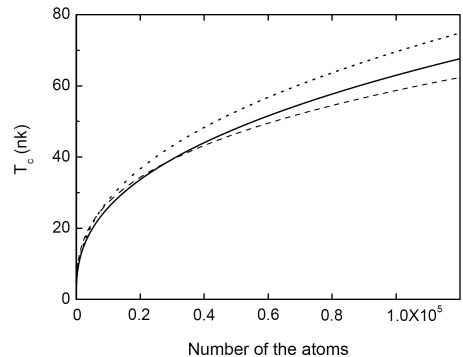}
\caption{Critical temperature $T_{c}$ versus the total number of
$^{87}\text{Rb}$ atoms. The solid curve and the dashed curve are obtained from
the numerical calculation of Eq. \eqref{N-nc-pz} and the analytical
approximation of $T_{c}$ (Eq. \eqref{Tc}), respectively. The dotted line gives
the full numerical result by diagonalizing the single-particle Hamiltonian.
The lattice parameters are $d=400\,\text{nm}$ and $s=11.2\,E_{r}$. The
trapping frequencies of the harmonic potential are $\omega_{\bot}=2\pi\times
83.7\,\text{Hz}$ and $\omega_{z}=2\pi\times 7.63\,\text{Hz}$, respectively.}
\label{Tc-N}
\end{figure}

We recall that an ideal Bose gas trapped in a 3D harmonic potential undergoes
the phase transition to Bose-Einstein condensation at a temperature
\cite{Tc-no lattice} $k_{B}T_{c}=0.94\hbar\overline{\omega}N^{1/3}$. Comparing
this expression with Eq. \eqref{Tc}, one can see that $T_{c}$ is changed by a
factor of $\left(m/m^{*}\right)^{1/6}$ due to the presence of the 1D lattice.
Since $m^{*}$ is always larger than $m$ \cite{Ho} under tight-binding
approximation, the combined trap $T_{c}$ is actually reduced compared to the
case without lattice. Note also that a homogeneous 3D lattice system has a
reduced $T_{c}$ as well \cite{Ho}, but with a reducing factor $\sqrt{m/m^{*}}$
instead.

We have calculated the critical temperature $T_{c}$ for a $^{87}$Rb gas in the
combined trap (see Fig. \ref{Tc-N}). The trap parameters are intentionally
chosen to match our experiment which will be described in the later section.
The numerically calculated $T_{c}$ is displayed by the solid curve, while the
dashed curve is the analytic $T_{c}$ calculated according to Eq.\eqref{Tc}.
The discrepancy between the two curves becomes larger as the atom number $N$
is increased, showing that the accuracy of the analytic estimation becomes
worse for larger $N$. We thus use only the numerically calculated $T_{c}$ in
the following computations.
\begin{figure}[h]
\centering
\includegraphics[width=0.7\columnwidth,angle=0]{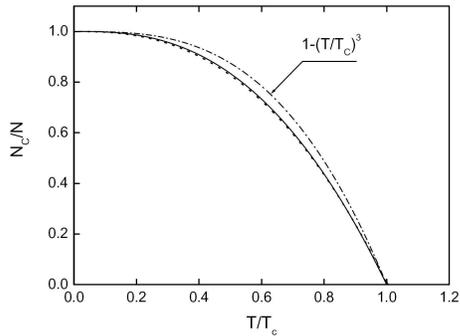}
\caption{Condensate fraction as a function of $T/T_{c}$. Solid line, numerical
results based on Eq. \eqref{N-nc-pz}. Dotted line is the full numerical result
based on diagonalizing the single-particle Hamiltonian. The dash dot line is
$1-(T/T_{c})^3$ for $T\leq T_{c}$. The parameters of the combined trap are the
same as in Fig. \ref{Tc-N}, and the atomic number is $N=5\times 10^{4}$,
corresponding to $T_{c}=47.9\,\text{nK}$.} \label{CondensateFraction}
\end{figure}

It is well known that an ideal Bose gas in a 3D harmonic potential shows a $T$
dependence of the condensate fraction as $N_{c}\sim1-(T/T_{c})^{3}$ for
$T<T_{c}$. We have also calculated the condensate fraction for our combined
trap system with $5\times 10^{4}$ atoms, as shown by the solid line in Fig.
\ref{CondensateFraction}. It displays a noticeable deviation from the curve of
$1-(T/T_{c})^{3}$, but fits well to the characteristic shape,
$1-(T/T_{c})^{\alpha}$, with $\alpha=2.679$.

To justify our analytical approximation, we also calculate the critical
temperature and condensate fraction based on the diagonalization of the
single-particle Hamiltonian. The energy spectrum of the system is written as
\begin{equation}
\varepsilon _{n_{x}n_{y}n_{z}}=\hbar \omega _{\perp }\left(
n_{x}+n_{y}+1\right) +\varepsilon _{n_{z}}.
\end{equation}
$\varepsilon _{n_{z}}$ can be obtained numerically from the
following single-particle Hamiltonian along the $z$ direction,
\begin{equation}
\widehat{H}_{z}=-\frac{J}{2}\sum_{\left\langle i,j\right\rangle }\left(
\widehat{a}_{i}^{\dag }\widehat{a}_{j}+\widehat{a}_{i}\widehat{a}_{j}^{\dag
}\right) +\sum_{i}\varepsilon _{i}\widehat{a}_{i}^{\dag }\widehat{a}_{i}.
\end{equation}
Here $\varepsilon _{i}$ describes an energy offset at each lattice
site due to the presence of the harmonic trap along the $z$
direction. By diagonalizing the matrix $\left\langle i\left\vert
\widehat{H}_{z}\right\vert j\right\rangle $, one can get directly
the energy spectrum $\varepsilon _{n_{z}}$. Furthermore, with the
following formula,
\begin{equation}
N=\sum_{n_{x},n_{y},n_{z}}\frac{1}{e^{\left( \varepsilon
_{n_{x}n_{y}n_{z}}-\mu \right) /k_{B}T}-1},
\end{equation}
we give the full numerical results of the critical temperature and
condensate fraction by the dotted lines in Figs. \ref{Tc-N} and
\ref{CondensateFraction}, respectively. Clearly, our semiclassical
treatment agrees with the full numerical method, and proves to be
reliable. Furthermore, it offers a convenient way to analyze the
spatial distribution of a confined atomic cloud, which in turn
simplifies the calculation of interference patterns.

\section{INTERFERENCE PEAKS}
\label{Peaks}

When the combined trap is suddenly shut off at the moment $t=0$, the Bose gas
starts to expand freely. After a time of flight $\tau$, the expanded
wavepackets initially localized in single lattice wells overlap with each
other, forming a 3D density distribution. In the following calculation, the
$x$ and $y$ dependence of the atomic density will be integrated out so as to
obtain a density profile along the $z$ direction only. This is convenient for
making a comparison with the experimental results. Usually, an absorption
image is used to record the column density profile of a released atomic cloud.
Supposing that the probe laser beam is applied along the direction of the $x$
axis, the density profile along $z$ can be easily obtained by integrating the
column density over the $y$ dimension.

\subsection{Normal gas}
\label{normal gas}

In the combined trap, normal gas atoms are distributed over the
transverse harmonic modes labeled by a positive quantum number
$q=n_{x}+n_{y}$. For a given $q$, there are $q+1$ degenerate states,
and we hereafter call them substates. The summation over $n_{x}$ and
$n_{y}$ in the previously mentioned equations is thus equivalent to
$\sum_{q}(q+1)\cdots$. From Eq. \eqref{N-nc}, one sees that the
substates belonging to the same $q$ number have identical
populations and spatial distribution along the $z$ direction. Due to
optical lattice potential, atoms in a single substate are further
distributed over the Bloch states with different quasimomentum
$p_{z}$ with $p_{z}/\hbar\in(-\pi/d,\pi/d)$. Each $p_{z}$ component
can be treated semiclassically where the influence of the optical
lattice is given by a quantum wave packet description, while the
influence of the harmonic trap along the $z$ direction is treated
semiclassically. In such a picture, the single-particle wave
function of a $p_{z}$ component at $t=0$ takes the following form:
\begin{equation}\label{psi-nc-t0}
\Psi_{p_{z}}^{q}(t=0)=\sum_{l}\alpha_{l}^{q}w(z-ld)\exp (ip_{z}z/\hbar),
\end{equation}
where $(\alpha_{l}^{q})^{2}$ denotes the probability for a particle roughly
located in the $l$th lattice site for the transverse harmonic mode $q$.

Equation \eqref{N-nc} shows that the atomic density of a substate
with $p_{z}$ has an envelope as
\begin{equation}\label{nk-z}
n(z)=\frac{\Delta p_{z}}{2\pi\hbar}\frac{1}{e^{\beta [q \hbar\omega_{\bot}+2
J(1-\cos(p_{z}d/\hbar))+\frac{1}{2}m\omega_{z}^{2}z^{2}]}-1},
\end{equation}
where $\Delta p_{z}$ denotes a small interval of $p_{z}$. The atom number in
the $l$th lattice site is then
\begin{equation}\label{nk-l}
n_{l} = \frac{d\Delta p_{z}}{2\pi\hbar}\frac{1}{e^{\beta [q
\hbar\omega_{\bot}+2
J(1-\cos(p_{z}d/\hbar))+\frac{1}{2}m\omega_{z}^{2}d^{2}l^{2}]}-1}.
\end{equation}
Therefore, $\alpha_{l}^{q}$ is simply given by
$(\alpha_{l}^{q})^{2}=n_{l}/N_{q}$, with $N_{q}=\sum_{l}n_{l}$ being the total
atom number of the $p_{z}$ component in the substate of interest. In
principle, $\alpha_{l}^{q}$ should be determined by solving the Schrodinger
equation of $H_{z}$. However, as shown lately, the thermal average of in-trap
density written in terms of $|\alpha_{l}^{p}|^{2}$ is matched to the
expression obtained by semiclassical approximation, hence within semiclassical
approximation $|\alpha_{l}^{q}|^{2}$ can be identified to
$(\alpha_{l}^{q})^{2}=n_{l}/N_{q}$.

In the tight-binding limit $w(z)$ can be well approximated by a Gaussian wave
packet $(\pi\sigma^{2})^{-1/4}\exp (-z^{2}/2\sigma^{2})$, where
$\sigma=\sqrt{\hbar/m\widetilde{\omega}_{z}}$ is the oscillator length. After
the free expansion over a time of $\tau$, the single-particle wave function of
the atoms with $p_{z}$ is written as
\begin{equation}\label{psi-nc-tau}
\begin{split}
& \Psi_{p_{z}}^{q}(t=\tau)\\
&=\sum_{l}\alpha_{l}^{q}\int K(z,z',\tau)w(z'-ld)\exp
(ip_{z}z'/\hbar)\text{d}z'.
\end{split}
\end{equation}
Here, $K(z,z',\tau)=\langle z|\exp(-iH\tau/\hbar)|z'\rangle$ is the
propagator, with $H$ the Hamiltonian governing the expansion process. If the
interatomic interaction is neglected, $H$ contains only the kinetic energy,
say, $H=P_{z}^2/2m$. In this case, it is straightforward to get
\begin{equation}\label{propagator}
K(z,z',\tau)=\sqrt{\frac{m}{i2\pi\hbar\tau}}\exp\left[
\frac{im}{2\hbar\tau}(z-z')^{2}\right].
\end{equation}

For simplicity of expression and calculation, we will use dimensionless units
for the length in $z$, the quasimomentum $p_{z}$ and the time $t$ by the
replacement $z\rightarrow zd$ (and hence $\sigma\rightarrow\sigma d$),
$p_{z}\rightarrow p_{z}\hbar/d$, and $t\rightarrow t(2md^{2}/\hbar)$.
Inserting Eq. \eqref{propagator} into Eq. \eqref{psi-nc-tau}, and working out
the integration over $z'$, one gets
\begin{equation}\label{psi-nc-tau-1}
\Psi_{p_{z}}^{q}(t=\tau)=A\sum_{l}\alpha_{l}^{q}B(p_{z},l,z),
\end{equation}
with $A$ and $B$ given by
\begin{equation*}\label{A-tau}
A=\pi^{-1/4}(\sigma+i2\tau/\sigma)^{-1/2},
\end{equation*}
\begin{equation*}\label{B-tau-l-z}
\begin{split}
B(p_{z},& l,z)=\\
& \exp\left[\frac{-2\tau\sigma^{2}p_{z}^{2}
+2\sigma^{2}zp_{z}+i[4lp_{z}\tau+(z-l)^{2}]}{2(\sigma^{4}+4\tau^{2})/(2\tau+i\sigma^{2})}\right].
\end{split}
\end{equation*}
\begin{figure}[h]
\centering
\includegraphics[width=0.7\columnwidth,angle=0]{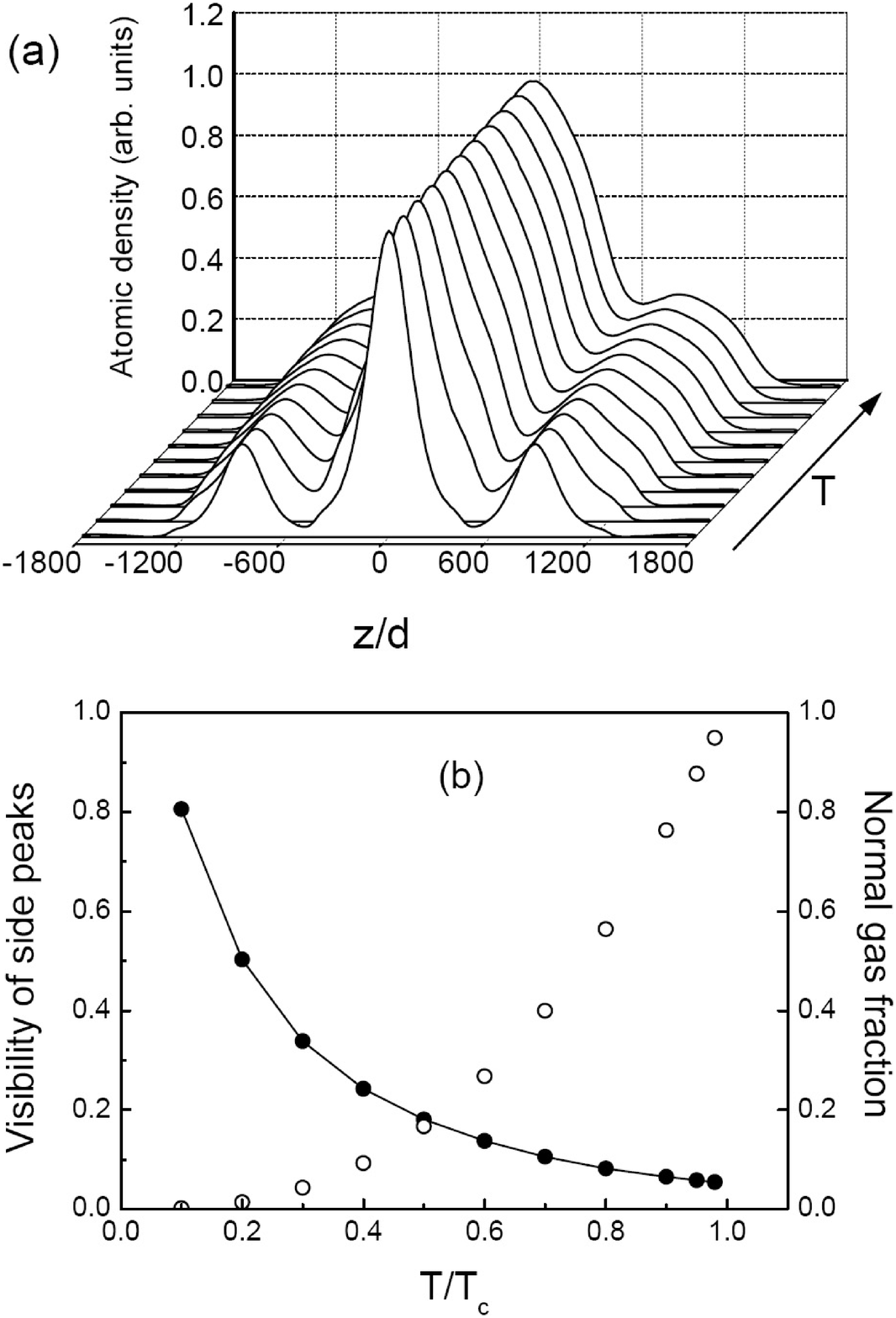}
\caption{(a) The solid lines are the calculated atomic distribution
of the normal gas ($^{87}$Rb) after $30\,\text{ms}$ of time of
flight. The total atom number $N=5.9\times 10^{4}$, and the trap
parameters are the same as in Fig. \ref{Tc-N}, corresponding to
$T_{c}=51.1\,\text{nK}$. For the top two curves, the temperature
$T/T_{c}$ is $0.98$ and $0.95$, respectively. The other nine curves,
from top to bottom, are for $T/T_{c}$ ranging from $0.9$ to $0.1$
with a step of $0.1$. Vertical scales of the curves have been
adjusted so that the central peaks have roughly the same height. (b)
Solid circles represent the visibility of the side peaks in (a). The
solid curve connecting the points is added to guide the eye. Open
circles show the normal gas fraction for the given total atom
number.} \label{Normalgaspeaks}
\end{figure}

Taking into consideration all transverse modes and all $p_{z}$
components, one gets the atom density after the time of flight,
\begin{equation}\label{n_normal-z}
\begin{split}
n_{nc}(z)&=\sum_{q=1}^{\infty}
\sum_{p_{z}}(q+1)N_{q}|\Psi_{p_{z}}^{q}(t=\tau)|^{2} \\
&=|A|^{2}\sum_{q=1}^{\infty}\sum_{p_{z}}(q+1)
\left|\sum_{l}\sqrt{n_{l}}B(p_{z},l,z)\right|^{2}.
\end{split}
\end{equation}
In the numerical calculations of Eq. \eqref{n_normal-z}, the summation over
transverse modes is cutoff at $q=200$, while $n$ is cutoff at $30$ and lattice
number $l$ at $\pm 350$. These cutoff numbers are chosen to assure a high
accuracy better than $0.2\%$ in the calculations of atom numbers. By setting
$\Delta p_{z}$ to $0.05\pi$, the whole $p_{z}$ range is divided into $40$
intervals. This step size of $p_{z}$ has the order of $h/dM$, where $M\simeq
100$ is the typical spatial extent in the $z$ direction. We have also checked
that the calculated results have almost no change when further reducing
$\Delta p_{z}$. The step size in $z$ is set to be $18d$ ($7.2\,\mu \text{m}$),
comparable to the pixel size of $9\,\mu \text{m}$ in our experiment.

In Fig. \ref{Normalgaspeaks}(a) we show the numerical results of a normal gas
of $^{87}$Rb atoms after $30\,\text{ms}$ of time of flight. It is obvious that
the normal gas leads to three peaks in the atomic distribution along the $z$
dimension. However, these peaks are not sharp, and, close to the $T_{c}$, the
side peaks are not even well resolved. In contrast, a normal gas initially
trapped in a 3D homogenous lattice system gives rise to much sharper peaks
\cite{Ho}.

We now define a visibility for the side peaks as in Ref. \cite{Lattice depth}:
\begin{equation}\label{Visibility}
v=\frac{n_{A}-n_{B}}{n_{A}-n_{B}},
\end{equation}
where $n_{A}$ is the atomic density of the side peak, and $n_{B}$ is atomic
density at the minimum between the central peak and the side peak. From Fig.
\ref{Normalgaspeaks}(b) it can be seen that the visibility $v$ is well below
$1$ at a considerable fraction of the normal gas. Although $v$ can reach $0.8$
at a very low temperature of $T=0.1T_{c}\thicksim 5\,\text{nK}$, only about
$0.1\%$ of the atoms remain in the normal gas state while all other atoms are
condensed. Actually, it is hard to detect such a small number of atoms using
the conventional absorption imaging method.

\subsection{Condensed gas}
\label{Condensed gas}

Unlike the normal gas, Bose-condensed atoms in the combined trap
occupy the lowest state of $q=0$ and pile to a small quasimomentum
interval of $p_{z}=0$. Nevertheless, the normal gas propagator holds
also for the condensed gas. In analogy with the calculations for the
normal gas, one can derive the single-particle wave function of the
condensed gas after the time of flight,
\begin{equation}\label{psi-n0-tau}
\Psi_{0}(t=\tau)=A(\tau)\sum_{k}\alpha_{k}B_{0}(k,z),
\end{equation}
where $k$ denotes the $k$th lattice well, $\alpha_{k}^{2}$ is the probability
of an atom staying in the $k$th well, and
\begin{equation*}\label{B0-tau-k-z}
B_{0}(k,z)=\exp\left[\frac{(i2\tau-\sigma^{2})
(z-k)^{2}}{2(\sigma^{4}+4\tau^{2})}\right].
\end{equation*}
Note that, in the two formulas above, $\tau$, $z$ and $\sigma$ are in their
dimensionless form.

In the tight-binding limit, condensed atoms in the combined trap form an array
of subcondensates along the $z$ axis. Each subcondensate is a 2D quantum gas
in nature, and its density distribution in the radial dimensions is described
by a Thomas-Fermi profile \cite{g2D-Pedri}. The local chemical potential
associated with the repulsive interaction of the atoms depends upon the
average atom number in the following form \cite{Miu-Anker}:
\begin{equation}\label{miuk}
\mu_{k}^{\text{loc}}=\sqrt{\frac{gm\omega_{\bot}^{2}N_{k}}{\sqrt{2}\pi^{3/2}\sigma}}.
\end{equation}
Here, $N_{k}$ is the average atom number in the $k$th lattice well,
and $g=4\pi\hbar^{2}a/m$ is the interaction parameter with $a$ the
$s$-wave scattering length. The sum of $\mu_{k}^{\text{loc}}$ and
the external harmonic potential $(1/2)m\omega_{z}^{2}z^{2}$ is just
the chemical potential which should remain invariant throughout the
entire condensed gas at equilibrium. Accordingly, $N_{k}$ is given
by
\begin{equation}\label{Nk}
N_{k}=(15N_{c}/16k_{M})(1-k^{2}/k_{M}^{2})^{2},
\end{equation}
where $k_{M}$ labels the outermost lattice well populated with condensed
atoms, and it is written as \cite{g2D-Pedri}
\begin{equation}\label{kM}
k_{M}^{2}=\frac{2\hbar\overline{\omega}}{m\omega_{z}^{2}d^{2}}
\left(\frac{15N_{c}}{8\sqrt{\pi}}\frac{a}{a_{ho}}\frac{d}{\sigma}\right)^{2/5}.
\end{equation}
If we neglect the mean-field interaction of the condensed gas during the free
expansion time, the density distribution after the TOF can be directly derived
from the coherent superposition of the expanded subcondensates:
\begin{equation}\label{n0-z}
n_{c}(z)=|A(\tau)|^{2}\left|\sum_{k=-k_{M}}^{k_{M}}\sqrt{N_{k}}B_{0}(k,z)\right|^{2}.
\end{equation}

\begin{figure}[h]
\centering
\includegraphics[width=0.7\columnwidth,angle=0]{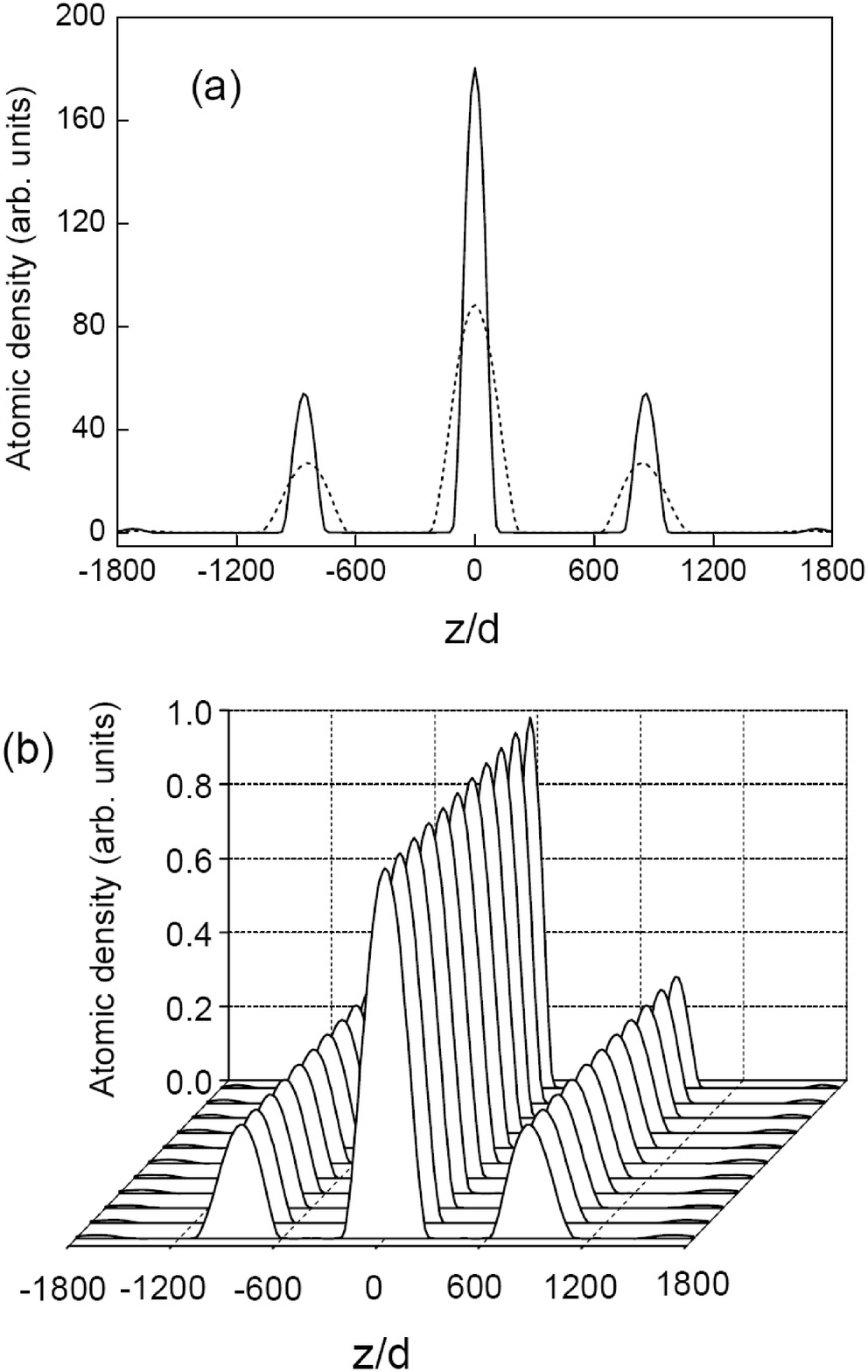}
\caption{Atomic distribution of the condensed gases after
$30\,\text{ms}$ of time of flight. The total atom number and the
trap parameters are the same as in Fig. \ref{Normalgaspeaks}. (a)
The dashed (solid) line is the result for $T=0.7T_{c}$ with
(without) a consideration of the mean-field interaction during TOF.
(b) These curves are calculated from Eq. \eqref{n0-z-mu}. From the
top curve to the bottom one, the temperature is decreased in
sequence, with the same values as in Fig. \ref{Normalgaspeaks}.
Vertical scales of the curves have been adjusted so that the central
peaks have the same height.} \label{Condensedgaspeaks}
\end{figure}

For the condensed gas before expansion, due to the high atomic
density, $\mu_{k}^{\text{loc}}/h$ is in the order of several hundred
Hertz. Although the atomic density drops quickly during the TOF, we
still expect that mean-field interaction might lead to a
considerable change in the coherence property of the expanding
atomic clouds. For simplicity, we only consider the mean-field
interaction within the single expanding subcondensates, and neglect
the interaction between them. At the beginning of the TOF, the
combined trap is suddenly turned off. Therefore, the total energy of
the $k$th subcondensate includes only the mean-field energy at this
moment, that is,
\begin{equation*}\label{E-k}
E_{k}=E_{\text{int}}=(1/2)U_{k}N_{k}^{2}.
\end{equation*}
Here, $U_{k}=g\int|\Phi_{k}(\textbf{r},z)|^{4}\text{d}z\text{d}\textbf{r}$ is
the on-site interaction matrix element of the $k$th subcondensate when
confined in the combined trap. Using the analytic form of
$\Phi_{k}(\textbf{r},z)$, we can get the expression of $U_{k}$ in terms of the
trap parameters:
\begin{equation}\label{U}
U_{k}=\frac{1}{3}\left(\frac{2}{\pi}\right)^{3/4}\sqrt{\frac{gm\omega_{\bot}^{2}}{\sigma
N_{k}}}.
\end{equation}
Its dependence on $N_{k}$ is due to the fact that the atomic number affects
the Thomas-Fermi radius of the radial wave-function. At later times, the total
energy $E_{k}$ remains constant despite the fact that the interaction energy
is being converted into kinetic energy. Then, the corresponding chemical
potential is simply given by $\mu_{k}=\partial E_{k}/\partial
N_{k}=\mu_{k}^{\text{loc}}/2$. Over the total time of flight, the $k$th
subcondensate acquires an additional phase factor $\exp(-i\mu_{k}\tau/\hbar)$.
Consequently, we can write the density distribution at the end of TOF by just
inserting this phase factor into Eq.\eqref{n0-z}:
\begin{equation}\label{n0-z-mu}
n_{c}(z)=|A(\tau)|^{2}\left|\sum_{k=-k_{M}}^{k_{M}}\sqrt{N_{k}}\exp(-i\mu_{k}\tau/\hbar)
B_{0}(k,z)\right|^{2}.
\end{equation}
From this equation one sees that $\mu_{k}$ will affect $n_{c}(z)$ by
its nonuniformity. In Fig. \ref{Condensedgaspeaks}(a), we plot two
curves calculated, respectively, with Eqs. \eqref{n0-z} and
\eqref{n0-z-mu} for a condensed gas of $2.5\times 10^{4}$ atoms.
When the mean-field interaction is included, all three interference
peaks are significantly broadened by about a factor of two. Since
the mean-field interaction is nonnegligible, all the theoretical
interference patterns mentioned hereafter are computed by Eq.
\eqref{n0-z-mu}.

In Fig. \ref{Condensedgaspeaks}(b), we plot the calculated density
distributions of the condensed gases at different temperatures for a
fixed total atomic number and a fixed lattice depth. With decreased
temperature, the condensate contains more atoms, leading to wider
interference peaks due to mean-field interactions. Unlike a normal
gas, a condensed gas always shows fully resolved interference peaks,
with a high visibility very close to $100\%$. This characteristic
behavior can be easily understood as the global coherence of
condensed atoms in a superfluid state. Additionally, these peaks are
considerably narrower than that of the corresponding normal gases,
except the extreme cases of very low temperatures that the
normal-gas atom number is very small and hardly detectable. When one
measures the interference pattern of a mixture of the condensed gas
and a normal gas, one would see three narrow peaks riding on three
broad peaks. This is the so-called ``peak on a peak" structure which
was first predicted for a homogeneous system \cite{Ho,Kato}. For an
inhomogeneous system in the combined tap, the onset of the condensed
gas is also characterized by the ``peak on a peak" structure.

On the other hand, if the condensed gas undergoes only a ballistic expansion
during the TOF (no mean-field interaction), the relative intensity of the side
peaks with respect to the central one should obey a simple law
\cite{g2D-Pedri}: $P_{1}=\exp(-4/\sqrt{s})$. We check the data in Fig.
\ref{Condensedgaspeaks}(b) ($P_{1}=0.303$ for $s=11.2$), and find that the
side peak intensities agree well with $P_{1}$ (to within $2$ percent). It
seems that the analytic expression of $P_{1}$ is also valid in the case of the
existence of mean-field interaction during the expansion time.

\section{EXPERIMENT}
\label{Experiment}

In experiment, we create a cigar-shaped $^{87}$Rb condensate in the
hyperfine state $|F=2, M_{F}=2\rangle$, confined in a conventional
Quadrupole Ioffe Configuration (QUIC) trap with an axial trapping
frequency of $\omega_{z}=2\pi\times18.7$ Hz and radial trapping
frequency of $\omega_{\bot}=2\pi\times205$ Hz. A nearly pure
condensate contains approximately $2\times10^{5}$ atoms. If the
frequency of the rf knife for evaporation cooling is ramped down
further, we can obtain a condensate with a lower temperature at the
cost of decreased atomic numbers. Certainly, the temperature is hard
to measure because there are almost no thermal atoms remained.
Nevertheless, we are able to coarsely adjust the temperature of the
cold atomic sample using the rf knife. After the evaporation
cooling, the QUIC trap is adiabatically relaxed until the axial and
radial trapping frequencies are lowered to $\omega_{\bot}=2\pi\times
83.7\,\text{Hz}$ and $\omega_{z}=2\pi\times 7.63\,\text{Hz}$,
respectively. Accordingly, the spatial extension of the condensate
wave packet is increased by a factor of $2.45$, so as to cover more
lattice periods at later times. The optical lattice is formed by one
retroreflected laser beam which is derived from a Ti:sapphire laser
at a wavelength of $\lambda=800\,\text{nm}$ and focused to a
$1/e^{2}$ radius of  $300\,\mu \text{m}$. It is applied to the
condensate along the long axis, and it is ramped up to a given
intensity over a time of $50\, \text{ms}$ and held at this value for
$10\, \text{ms}$. The sum of the optical lattice and the QUIC trap
potential gives a combined trap in accord with Eq. \eqref{Combined
trap}. The potential depth of the optical lattice is calibrated
using the method of Kapitza-Dirac scattering \cite{Lattice depth}.
We then suddenly switch off the combined trap and allow the cold
atomic sample to expand freely for a time of $30\,\text{ms}$.
Finally, we take an absorption image of the expanded atomic gas
using a CCD camera, from which we can deduce both the total atom
number and the atomic density distribution.
\begin{figure}[h]
\centering
\includegraphics[width=0.7\columnwidth,angle=0]{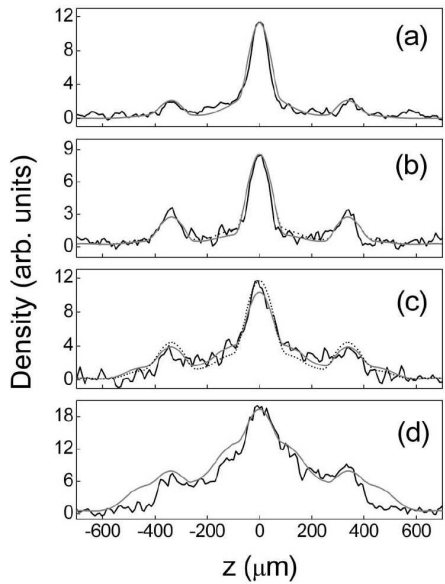}
\caption{Density distribution of two typical atomic samples of
$^{87}$Rb. Black curves are the measured linear density of the
released $^{87}$Rb atoms after a $30\,ms$ TOF. Gray curves are the
calculated results based on Eqs. \eqref{n_normal-z} and
\eqref{n0-z-mu}, where temperature is used as a fitting parameter.
The dotted curves in (b) and (c) are the calculated results based on
measured temperature values.  Their vertical scales have been
adjusted for comparison with the black curves. (a) The measured
total atom number and lattice depth are $N\simeq5.3\times 10^{4}$
and $s\simeq5.6$, respectively, corresponding to
$T_{c}=55.9\,\text{nK}$. $T=44.7\,\text{nK}$ is assumed in the
computation. (b) $N\simeq5.9\times 10^{4}$, $s\simeq11.2$ and
$T_{c}=51.1\,\text{nK}$; $T=33\,\text{nK}$ for the gray line and
$T=37.4\,\text{nK}$ for the dotted line. (c) $N\simeq1.1\times
10^{5}$, $s\simeq16.7$, and $T_{c}=63.0\,\text{nK}$;
$T=55\,\text{nK}$ for the gray line and $T=49.3\,\text{nK}$ for the
dotted line. (d) $N\simeq 2\times 10^{5}$, $s\simeq 20$, and
$T_{c}=76.5\,\text{nK}$; $T=73.4\,\text{nK}$ for the gray line. }
\label{MeasuredPattern}
\end{figure}

The ``peak on a peak" features of interference patterns were
confirmed by the measured linear densities of expanded atomic clouds
in many runs of experiments. The black curves in Figs.
\ref{MeasuredPattern}(a)-\ref{MeasuredPattern}(c) display three
typical density distributions which were obtained by integrating the
pixels in each column of the corresponding absorption image. The
calculated critical temperature $T_{c}$ is usually in the order of
several tens of $\text{nK}$. In contrast, during the evaporative
cooling stage, the critical temperature for condensation in the QUIC
trap is much higher ($\sim400\,\text{nK}$). Actually, the experiment
reported in \cite{Burger} showed clearly a significantly decreased
critical temperature for a combined trap when compared with a purely
magnetic trap.

\begin{figure}[h]
\centering
\includegraphics[width=0.7\columnwidth,angle=0]{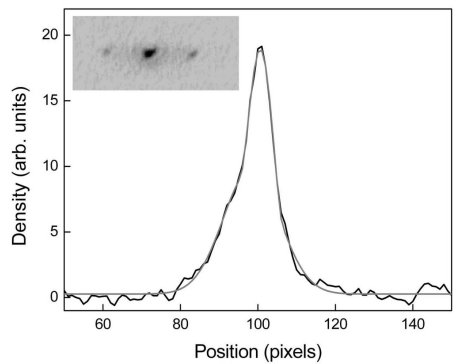}
\caption{Density distributions in the transverse dimension for the
same atomic cloud as in Fig. \ref{MeasuredPattern}(b). The black
curve is the integration along the axial direction of the central
peak. The gray curve is a bimodal fitting of the black curve, which
represents a superposition of a Gaussian profile and a Thomas-Fermi
profile (inverted parabola). The rms size of the Gaussian part gives
a temperature of $37.4\,\text{nK}$. (Inset) The absorption image of
this atomic cloud.} \label{Transverse}
\end{figure}

To test our theory, the temperature of the atomic sample before
expansion must be known. It can be deduced from the transverse
distribution of the normal gas after the TOF, which should take a
Gaussian profile due to the initially thermal occupation of the
transverse modes. Since the condensate part should take a
Thomas-Fermi profile in the transverse direction, a bimodal
transverse distribution is expected for a released gas, just as
shown in Fig. \ref{Transverse}. However, a measured temperature
based on this method usually has a large uncertainty due to the
following reasons. First, in the transverse direction, the spatial
extent of the condensate is not very distinct from that of the
normal gas, especially when the condensate fraction is large.
Second, the normal gas density profile deviates from an ideal
Gaussian shape, and exhibits a slight asymmetry that may arise from
the misalignment between the lattice light and the magnetic trap.
Third, the optical noise in the absorption images also lowers the
fitting accuracy. As pointed out in \cite{T-vs-entropy}, an atomic
sample can be significantly heated or cooled when adiabatically
loaded to an optical lattice. Yet, to date, we have no alternative
methods for accurate measurement of the temperature of an atomic
sample confined in a lattice system. We have to treat temperature as
a fitting parameter in the calculation, so that the calculated
density distribution most closely reproduces the experimental curve.

The gray curves in Fig. \ref{MeasuredPattern} are the calculated
density distributions of cold rubidium gases. Figure
\ref{MeasuredPattern}(a) is a case with a smaller atom number, and
the fitting curve agrees fairly well with the experimental data.
Figure \ref{MeasuredPattern}(b) displays the interference pattern of
another atomic sample initially confined in a deeper lattice, and
the side peaks are more prominent. As a comparison, the atomic
sample in Fig. \ref{MeasuredPattern}(c) contains more atoms and the
lattice depth was further increased. Accordingly, the calculated
$T_{c}$ shifts up to $63\,\text{nK}$. Since the normal gas density
is increased, the feature of ``peak on a peak" is more pronounced.
In (b) and (c), we also plot the density distributions calculated
using the measured temperature values. The larger deviation from the
measured density profiles should be attributed to the inaccuracy of
temperature.

Despite the overall agreement between the theoretical and
experimental curves in Fig. \ref{MeasuredPattern}, there are still
noticeable discrepancies. As the lattice depth increases,
theoretical normal gas peaks become broader than the measured
density profiles. As shown in Fig. \ref{MeasuredPattern}(d), the
gray line does not match the black curve, particularly at the wings
of the normal gas. For this atomic sample, the temperature is close
to $T_{c}$, and the condensate peaks are hence very small. The
mismatch between the theory and experiment indicates that our model
is not valid for very deep optical lattices. This can be easily
understood by comparing the tunneling energy $J$ to the energy
offset between adjacent lattice sites induced by harmonic-potential.
For a cloud with an extension of $l_{M}d$, this energy offset is
$m\omega_{z}^{2}d^{2}l_{M}$ at the site of $l_{M}$. A typical value
of $l_{M}=200$ corresponds to an energy offset of $2\pi\hbar\times
16\,\text{Hz}$, whereas $J$ decreases with increased lattice depth.
For $s=20$ as in Fig. \ref{MeasuredPattern}(d), $J\simeq
2\pi\hbar\times 10\,\text{Hz}$. When $J$ gets smaller than the
energy offset between adjacent lattice sites, normal gas atoms are
essentially localized and the semiclassical analysis breaks down. We
have also calculated the $T_{c}$'s for situations of $s\geq20$,
using the diagonalization method and semiclassical approximation,
respectively. We do find significant discrepancy between the
predictions of these two methods. In such situations, an atomic
cloud should be treated as a chain of separate condensates, where
the loss of condensate is interpreted as the loss of well-to-well
phase coherence \cite{Lattice depth}. At a depth level of
$s\simeq30$, we observed a completely disappearance of interference
peaks.

On the other hand, in our model, the interatomic interactions during
TOF are taken into account for the subcondensates individually. This
amounts to neglecting the variation of the wavefunction modulus
induced by the interactions between subcondensates. The fluctuations
of the wavefunction modulus are related to the relative phase of the
subcondensates, and hence affect their phase coherence, leading to
variations of the interference peak of the condensates. Evidently,
the calculated condensate peaks are slightly wider than the measured
ones (see Fig. \ref{MeasuredPattern}). Perhaps, the neglected
interactions are favorable for establishing a uniform phase which
partially cancels the phase nonuniformity discussed in Sec.
\ref{Condensed gas}. Since we have not found a simple model to
account for it, this effect will not be discussed in detail in this
paper.

\section{CONCLUSION}
\label{Conclusion}

We have performed a study, both theoretically and experimentally, on
the phase transition to macroscopic superfluidity for a Bose gas
confined in a combined trap formed by a harmonic potential and an
optical lattice. We have mainly investigated the interference
patterns of the Bose gases below the critical temperature. By using
a semiclassical energy spectrum and tight-binding approximation, we
have derived an analytical approximation of the critical temperature
which is applicable to an atomic gas residing in the vicinity of the
bottom of the ground Bloch band. Furthermore, the interference
patterns of the normal gas and the condensed gas were analyzed
separately. We have derived the analytical expressions of the atomic
density distribution for the released normal gas and condensed gas
which has experienced a free expansion over a time of flight. Our
calculation results show that a condensed gas is characterized by
fully-resolved narrow interference peaks while a normal gas forms
broad interference peaks with lower visibility. For comparison, we
have performed a preliminary experiment using Bose-Einstein
condensates of $^{87}$Rb atoms. The combined trap system was
implemented by applying a 1D optical lattice to a magnetically
trapped condensate. The measured interference pattern agrees
essentially with our theoretical prediction, exhibiting ``peak on a
peak" structures associated with the onset of condensed gases. Thus,
both the theoretical and experimental results confirm that the
``peak on a peak" structure is indeed a signature of macroscopic
superfluidity in our inhomogeneous lattice system.

\begin{acknowledgments}
We appreciate the helpful discussions with Xiaoji Zhou. This work is
supported by the National Natural Science Foundation of China under
Grant No. 10574142, and by the National Key Basic Research and
Development Program of China under Grant No. 2006CB921406 and No.
2011CB921503.

\end{acknowledgments}

\end{document}